\documentstyle{article}

\def\={\discretionary{-}{}{-}}
\def\ds{\displaystyle}

\def\sep#1{\setlength{\arraycolsep}{#1\arraycolsep}}

\def\astretch#1{\renewcommand{\arraystretch}{#1}}

\def\cako{Cauchy-Kowalevsky Theorem}
\def\caku{Cartan-Kuranishi Theorem}

\def\cc{Cartan character}

\def\de{differential equation}

\def\dreg{$\delta$-regular}
\def\Dreg{$\delta$-regularity}

\def\hp{Hilbert polynomial}
\def\intc{integrability condition}

\def\ode{ordinary \de}
\def\pde{partial \de}
\def\pss{power series solution}
\def\tc{Taylor coefficient}
\def\aam{a^\alpha_\mu}
\def\aq#1#2{\alpha_{#1}^{(#2)}}
\def\bq#1#2{\beta_{#1}^{(#2)}}

\def\dalembert{\raisebox{-1pt}{$\Box$}}

\def\diff#1#2{\frac{\partial#1}{\partial#2}}
\def\dim{{\rm dim}\,}
\def\E{{\cal E}}

\def\je#1{J_{#1}\E}

\def\Mc#1{{\cal M}_{#1}}
\def\Mcp#1#2{{\cal M}_{#1}^{(#2)}}
\def\Mcpp#1#2#3{\Mcp{#1+#2}{#3}}
\def\pam{p^\alpha_\mu}
\def\phit{\Phi^\tau}

\def\Rc#1{{\cal R}_{#1}}
\def\Rcp#1#2{{\cal R}_{#1}^{(#2)}}
\def\Rcpp#1#2#3{\Rcp{#1+#2}{#3}}
\def\rank{{\rm rank}\,}
\def\restrict#1#2{\setlength{\rulen}{#1pt}
                  \setlength{\ruoff}{0.5\rulen}\addtolength{\ruoff}{-2pt}
                  \setlength{\rebox}{\ruoff}\addtolength{\rebox}{-2pt}
                  \rule[-\ruoff]{0.3pt}{\rulen}
                  \raisebox{-\rebox}{$\hspace{2pt}_{#2}$}}
\def\s#1#2#3{\mbox{$s^{(#1)}_{#2}(#3)$}}

\newlength{\rulen}     
\newlength{\ruoff}     
\newlength{\rebox}     
\newlength{\boxdim}
\def\boxit#1{\framebox[\textwidth]%
                  {\begin{minipage}{\boxdim}%
                   {\renewcommand{\baselinestretch}{1}\small%
                    \vspace*{\baselineskip}#1\vspace*{\baselineskip}}%
                   \end{minipage}}}
\newtheorem{theorem}{Theorem}
\newtheorem{definition}[theorem]{Definition}

\begin{document}

\title{Involution and Constrained Dynamics~I: The Dirac Approach}
\author{Werner M. Seiler\thanks{On leave from Institut f\"ur Algorithmen
          und Kognitive Systeme, Universit\"at Karls\-ruhe, 76128 Karlsruhe,
          Germany},
        Robin W. Tucker\\
           School of Physics and Materials\\
           Lancaster University\\
           Bailrigg, LA1 4YB, UK\\
           Email: W.Seiler@lancaster.ac.uk, R.W.Tucker@lancaster.ac.uk}
\date{}
\maketitle

\begin{center}
\bf Abstract
\end{center}
\begin{quote}
{\small We study the theory of systems with constraints from the point of view
of the formal theory of partial differential equations. For finite-dimensional
systems we show that the Dirac algorithm completes the equations of motion to
an
involutive system. We discuss the implications of this identification for field
theories and argue that the involution analysis is more general and flexible
than the Dirac approach. We also derive intrinsic expressions for the number of
degrees of freedom.}
\end{quote}

\section{Introduction}

Constrained dynamics represents a cornerstone of theoretical physics, as every
relativistic theory and every theory with gauge symmetries necessarily
possesses
constraints. Thus it is not very surprising that many methods for dealing with
such systems have been developed (see e.g.\
\cite{hrt:ham,ht:quant,lus:noe,sm:dyn,sun:condyn}).  The purpose of this and
the
following articles in this series is to present an alternative ansatz based on
the modern theory of \de s and especially on the concept of
involution~\cite{pom:eins,wms:diss}.

The classical Hamiltonian treatment of systems with constraints was developed
by
Dirac~\cite{dir:ghd1,dir:ghd2,dir:qm}.  We will show that in the case of
finite-dimensional systems his algorithm corresponds to rendering the equations
of motion involutive. In the language of exterior differential systems this was
already noted by Hartley {\em et al.\/}~\cite{htt:constr}. But we will also
show
that this connection does no longer hold for field theories. Here it might
happen that the Dirac analysis alone is not sufficient to obtain all
constraints.

This identification appears natural, as the basic idea behind the Dirac
algorithm is to check whether or not the equations of motions are consistent.
But the notion of involution represents essentially a mathematical formulation
of this problem.  We believe that this approach has conceptual and practical
advantages, especially in the case of field theories. The theory of involution
is well understood for arbitrary systems. The \caku\ \cite{kur:pro,pom:eins}
yields a general procedure to complete any system of \pde s to an involutive
one.

In contrast one can find in physics many different approaches depending on
whether one deals with a system in Lagrangian or Hamiltonian formulation, or
whether the Lagrangian contains higher order derivatives or is linear in the
velocities. Each case is handled individually in the literature. A closer
analysis shows, however, that almost all proposed methods represent nothing
else
than special cases of the general completion procedure stemming from the
\caku\footnote{This holds even for the approach of Lusanna~\cite{lus:noe} via
the Noether theorem, as it is based on the eigenvectors of the Hessian. But
these correspond exactly to the linear combinations of the equations of motion
which yield the \intc s.}.

We study in this article the standard situation of a system described by a
Lagrangian depending on the generalized coordinates and the velocities as well
as Lagrangians containing higher order derivatives. The special case of a
Lagrangian being linear in the velocities and the so-called symplectic
formalism
of Faddeev and Jackiw~\cite{fj:ham} will be considered in the next article in
this series.  As concrete examples we consider among others the rigid rotator
and Podolsky's generalized electrodynamics.

Special emphasis is put on the problem of determining the number of degrees of
freedom. We propose a new intrinsic definition for field theories using the
\cc s of the field equations. It can also handle systems described in
characteristic coordinates like light-cone coordinates. The classical approach
based on a distinction into first and second class constraints fails in such a
situation~\cite{stein:probl}, as seemingly too many constraints occur.

The article is organized as follows: The next two sections serve as a brief
introduction into the formal theory of \de. They define the notion of
involution
and show how one completes an arbitrary system to an involutive one.
Section~\ref{dirac} reviews the classical Dirac approach. In
Section~\ref{point}
we show its relation to the formal analysis of the Hamiltonian equations of
motion for finite-dimensional systems. A detailed example is considered in
Section~\ref{ex}. Section~\ref{field} contains an example of a field theory
where the Dirac algorithm alone does not suffice to exhibit the full constraint
structure. The problem of counting degrees of freedom is tackled in
Sections~\ref{free} and~\ref{freeF} for the finite and infinite-dimensional
case, respectively. Finally, before some conclusions are given, we consider in
Section~\ref{hol} Lagrangians depending on higher order derivatives.

\section{Involution}\label{inv}

Formal theory uses a geometric approach to differential equations based on the
jet bundle formalism. It is beyond the scope of this paper to give a detailed
introduction into the underlying theory. The interested reader is referred to
the literature~\cite{pom:eins,sau:jet,wms:diss}. Here we are concerned with two
topics: the definition of an involutive system and how to compute the
arbitrariness of the general solution of such a system.

We will always work in a local coordinate system, although the whole theory can
be expressed in a coordinate free way. Let $X$ denote the space of the
independent variables $x_1,\ldots,x_n$ and let the dependent variables
$u^1,\ldots,u^m$ be fiber coordinates for the bundle~$\E$ over the base
space~$X$.  Derivatives are written in multi-index notation $\pam =
\partial^{|\mu|}u^\alpha /\partial x_1^{\mu_1}\cdots\partial x_n^{\mu_n}$ where
$|\mu|=\mu_1+\cdots+\mu_n$ is the length of the multi-index
$\mu=[\mu_1,\dots,\mu_n]$. Adding the derivatives~$\pam$ up to order~$q$
defines
a local coordinate system for the $q$-th order jet bundle~$\je{q}$.  A system
of
differential equation~$\Rc{q}$ of order~$q$ can be described locally by
\begin{equation}\label{sys}
\Rc{q}\,:\,\Bigl\{\ \Phi^\tau\left(x_i,u^\alpha,p^\alpha_\mu\right)=0\,,
\quad\tau=1,\dots,p\,;\ |\mu|\leq q\,.
\end{equation}
Geometrically, this represents a fibered submanifold of~$\je{q}$.

At least some of the ideas behind the concept of involution can be understood
best by considering the order by order construction of a formal power series
solution. For this purpose, we introduce the symbol~$\Mc{q}$ of a \de~$\Rc{q}$.

\begin{definition}\label{symbol}
The {\em symbol\/}~$\Mc{q}$ of the system~(\ref{sys}) is the solution space of
the following linear system of (algebraic!) equations in the
unknowns~$v^\alpha_\mu$
\begin{equation}\label{symbeq}
\Mc{q}\,:\,\left\{\sum_{\alpha,|\mu|=q}
\left({\partial\phit\over\partial\pam}\right) v^\alpha_\mu=0\,.\right.
\end{equation}
(By abuse of language, we will refer to both the linear system and its
solution
space as the symbol).
\end{definition}

The placeholders $v^\alpha_\mu$ are coordinates of a finite-dimensional vector
space, i.e.\ we introduce one coordinate for each derivative of order~$q$.
Definition~\ref{symbol} is most easily understood by considering a quasi-linear
system, i.e.\ a system linear in the derivatives~$\pam$ with $|\mu|=q$.  For
such a system the symbol is simply obtained by taking only the linear highest
order part and substituting $v^\alpha_\mu$ for~$\pam$.

We make a power series ansatz for the general solution of the differential
equation~$\Rc{q}$ by expanding around some point~$x^0$
\begin{equation}
u^\alpha(x)=\sum_{|\mu|=0}^\infty{\aam\over\mu!}(x-x^0)^\mu
\end{equation}
and substitute this ansatz into the equations~(\ref{sys}) evaluating
at~$x^0\,.$
This yields a system of algebraic equations for the Taylor coefficients~$\aam$
up to order~$q$.

The remaining coefficients can be computed by linear algebra only. For the
coefficients of order~$q+r$ we use the {\em prolonged\/} systems~$\Rc{q+r}$
which are obtained by differentiating each equation in~$\Rc{q}$ $r$~times
totally with respect to all independent variables. They are all quasi-linear.
If
we substitute again the power series ansatz into the prolonged
system~$\Rc{q+r}$
and evaluate at~$x^0\,,$ we get an inhomogeneous linear system for the
coefficients of order~$q+r$. Its homogeneous part is determined by the
prolonged
symbol~$\Mc{q+r}$, i.e.\ the symbol of~$\Rc{q+r}$.

The \tc s~$\aam$ of lower order appear in the matrix and in the right hand side
of this linear system. Thus we are able to express the coefficients of
order~$q+r$ through the coefficients of lower order. This is the precise
meaning
of constructing a power series order by order.

This construction will fail, if non-trivial \intc s occur, i.e.\ equations of
order~$q+r$ which are functionally independent of the equations contained in
the
prolonged system~$\Rc{q+r}$ and which are satisfied by every solution of the
system. Such equations arise usually by cross-differentiating and are detected
only in some higher prolongation. They pose additional conditions on the
coefficients of order~$q+r$.  Hence they must all be known to pursue the above
described procedure. We call a system which contains all its \intc s a {\em
formally integrable\/} system.

For formally integrable systems it is thus possible to construct order by order
a formal power series solution. The arbitrariness of the general solution is
reflected by the dimensions of the prolonged symbols, because at each order
$\dim\Mc{q+r}$ coefficients are not determined by the differential equations
but
can be chosen freely~\cite{wms:str}. Formal integrability does, however, not
suffice to determine these dimensions in advance without explicitly
constructing
the prolonged symbols.  This leads to the concept of involution.

We introduce the {\em class\/} of a multi-index $\mu=[\mu_1,\dots,\mu_n]\,.$ It
is the smallest~$k$ for which $\mu_k$ is different from zero. If we consider
the
symbol~(\ref{symbeq}) as a matrix, then its columns are labeled by the
coordinates~$v^\alpha_\mu\,.$ We order them by class, i.e.\ we always take a
column with a multi-index of higher class left of one with lower class. Then we
compute a row echelon form.

In this solved form the symbol is especially easy to analyze. Since we only
need
linear operations to obtain it, we can always perform the same operations with
the full system~$\Rc{q}$ and thus assume that (\ref{symbeq}) yields the symbol
directly in solved form. We denote the number of rows where the leading entry
or
pivot is of class~$k$ by~$\bq{q}{k}$ and we associate with each such row its
{\em multiplicative variables\/} $x_1,\dots,x_k\,.$

It is important to note that if we prolong each equation only with respect to
its multiplicative variables, we obtain independent equations, because each
equation will have a different leading term. The question is, whether
prolongation with respect to the non-multiplicative variables leads to
additional independent equations.  If not we call the symbol involutive.

\begin{definition}\label{invol}
The symbol~$\Mc{q}$ is called {\em involutive}, if
\begin{equation}\label{invsymb}
\rank\Mc{q+1}=\sum_{k=1}^n\,k\bq{q}{k}\,.
\end{equation}
The system~$\Rc{q}$ is called {\em involutive}, if it is formally integrable
and
its symbol is involutive.
\end{definition}

The above definition of the $\bq{q}{k}$ is obviously coordinate dependent. Thus
it seems, as if the involution of a symbol depends on the chosen coordinate
system, too. One can, however, show that almost every coordinate system leads
to
the same values for the $\bq{q}{k}$. These values are characterized by the
property that all the sums $\sum_{i=k}^n\bq{q}{i}$, $k=1,\dots,n$, are
maximal.\footnote{Note that this is different from requiring that the
$\bq{q}{k}$
themselves take maximal values!} A coordinate system which leads to these
values is called {\em $\delta$-regular}.  Definition~\ref{invol} assumes that
the $\bq{q}{k}$ are computed in such a coordinate system. Besides there exist
alternative methods to obtain the correct values
intrinsically~\cite{pom:eins,wms:tab,wms:diss}. We will return to this
point in Section~\ref{freeF}.

The prolongation of an involutive symbol is again involutive.  Since prolonging
an equation with respect to one of its multiplicative variables~$x_i$ yields an
equation of class~$i$, we get $\bq{q+1}{i}=\sum_{k=i}^n\bq{q}{k}\,.$ Inductive
use of this relation leads to
\begin{equation}
\bq{q+r}{k}=\sum_{i=k}^n\,{r+i-k-1\choose r-1}\bq{q}{i}
\end{equation}
and together with Definition~\ref{invol} to
\begin{equation}\label{rankM}
\rank\Mc{q+r}=\sum_{k=1}^n\,{r+k-1\choose r}\bq{q}{k}\,.
\end{equation}

Besides the possibility to predict the number of arbitrary Taylor coefficients
at any order, involutive systems have another advantage compared with formally
integrable ones. There exists an easily applicable criterion to check whether
or
not a system is involutive. The problem of the definition of formal
integrability is that one has to prove that a system does not generate
non-trivial \intc s at any prolongation order, i.e.\ one must check an infinite
number of conditions. This can, however, be done in a finite manner for systems
with an involutive symbol.

\begin{theorem}\label{crit}
Let $\Rc{q}$ be a $q$-th order differential equation with an involutive
symbol~$\Mc{q}$. If no \intc s arise during the prolongation of~$\Rc{q}$
to~$\Rc{q+1}$, then $\Rc{q}$ is involutive.
\end{theorem}

\section{Completion to Involution and Arbitrariness}\label{compl}

Since we have seen that involutive systems have many advantages, the question
naturally arises whether they form only a very special class of systems and
what
to do with a non-involutive system. The interesting answer is given by the
Cartan\=Kuranishi Theorem~\cite{kur:pro,pom:eins,wms:diss}.

\begin{theorem}
Any system~$\Rc{q}$ can be completed to an equivalent involutive one by a
finite
number of prolongations and projections (i.e.\ addition of \intc s).
\end{theorem}

Since this theorem depends on some fairly deep results in the formal theory, we
will not present a proof but only discuss an algorithm to perform this
completion. It is based on Theorem~\ref{crit} above and consists essentially of
two nested loops. The inner loop prolongs the system until its symbol becomes
involutive. The outer loop checks then for \intc s and adds them.  The
difficult
part of the proof is to show the termination of the inner loop.  The
termination
of the outer one follows from a simple Noetherian argument.

Involution of a symbol can be checked easily using Definition~\ref{invol}, if
we
assume that the coordinate system is $\delta$-regular what we will do in the
sequel. It requires only linear algebra. Whether or not \intc s arise during a
prolongation, can be deduced from a dimensional argument.

Denote the projection of the system~$\Rc{q+1}$ into the $q$-th order jet
bundle~$\je{q}$ by~$\Rcp{q}{1}$. Its dimension can be computed indirectly from
the identity
\begin{equation}\label{dim}
\dim\Rcp{q}{1}=\dim\Rc{q+1}-\dim\Mc{q+1}
\end{equation}
which reflects the fact that \intc s are connected with rank defects in the
symbol.  None has occurred during the prolongation from $\Rc{q}$ to~$\Rc{q+1}$,
if and only if this dimension is equal to~$\dim\Rc{q}$.

There are essentially two possible reasons for \intc s.  The classical one is
that it is possible by some linear combination of equations of order~$q+1$
in~$\Rc{q+1}$ to eliminate all derivatives of that order. This is a
generalization of the usual cross-derivative. The other one is that $\Rc{q}$
contains some equations of lower order. In order to construct $\Rc{q+1}$ all
equations in $\Rc{q}$ must be prolonged. If now some equations are of lower
order, it might happen that their prolongation leads to new independent
equations of order less than or equal to~$q$. They must be taken into account
in
the projection to~$\Rcp{q}{1}$.

Fig.~\ref{ckad} shows this algorithm in a more formal language.
$\Rcpp{q}{r}{s}$~denotes here the system obtained after $r+s$~prolongations and
$s$~projections. $\Mcpp{q}{r}{s}$~is the corresponding symbol. In this form it
is comparatively straightforward to implement it in a computer program. The
determination of the dimensions of the various submanifolds~$\Rcpp{q}{r}{s}$
poses the main remaining problem, especially for non-linear systems.
Ref.~\cite{wms:aci,wms:diss} describe an implementation in the computer algebra
system~{\small AXIOM}.

\begin{figure}
\boxit{\begin{tabbing}
[5.5.5]\quad \= \qquad \= \qquad \= \hspace{5 true cm} \= \kill
{}[1] \> r $\leftarrow$ 0; s$\leftarrow$ 0 \\
{}[2] \> compute $\Rc{q+1}$ \> \> \> {\em \{prolong\/\}}\\
{}[3] \> compute $\Mc{q},\Mc{q+1}$ \> \> \> {\em \{extract symbols\/\}}\\
{}[4] \> {\tt until} $\Rcp{q+r}{s}$ involutive {\tt repeat}\\
{}[4.1] \> \> {\tt while} $\#multVar(\Mcp{q+r}{s})\neq
                                 \rank\Mcp{q+r+1}{s}$ {\tt repeat}\\
{}[4.1.1] \> \> \> r $\leftarrow$ r+1
                \> {\em \{counter for prolongations\/\}}\\
{}[4.1.2] \> \> \> compute $\Rcp{q+r+1}{s}$ \> {\em \{prolong\/\}}\\
{}[4.1.3] \> \> \> compute $\Mcp{q+r+1}{s}$ \> {\em \{extract symbol\/\}}\\
{}[4.2] \> \> {\tt if} $\dim\Rcp{q+r+1}{s}-\dim\Mcp{q+r+1}{s}<
                                           \dim\Rcp{q+r}{s}$ {\tt then}\\
{}[4.2.1] \> \> \> s $\leftarrow$ s+1 \> {\em \{counter for projections\/\}}\\
{}[4.2.2] \> \> \> compute $\Rcp{q+r\phantom{+1}}{s}$
                   \> {\em \{add \intc s\/\}}\\
{}[4.2.3] \> \> \> compute $\Rcp{q+r+1}{s}$ \> {\em \{prolong\/\}}\\
{}[4.2.4] \> \> \> compute $\Mcp{q+r}{s},\Mcp{q+r+1}{s}$
          \> {\em \{extract symbols\/\}}\\
{}[5] \> {\tt return} $\Rcpp{q}{r}{s}$
\end{tabbing}}
\caption{Algorithm for the \caku\label{ckad}}
\end{figure}

For \ode s this algorithm becomes very simple. Since there is only one
independent variable, we find always an involutive symbol and cross-derivatives
are of course not possible. The only possibility for \intc s is the
prolongation
of lower order equations. For \pde s we recall that the other \intc s can
always
be found by considering the prolongations with respect to non-multiplicative
variables.

To conclude this section we briefly recall some results of Ref.~\cite{wms:str}
concerning the arbitrariness of the general solution which will be needed
later.
(\ref{rankM})~yields only the rank of the prolonged symbols, but their
dimensions are more interesting. They can be expressed in a similar way, if we
introduce the {\em\cc s\/}~$\aq{q}{k}$ of a \de
\begin{equation}\label{carchar}
\aq{q}{k}=m{q+n-k-1\choose q-1}-\bq{q}{k},\qquad k=1,\dots,n\,.
\end{equation}
They form a descending sequence
\begin{equation}\label{aseq}
\aq{q}{1}\geq\aq{q}{2}\geq\cdots\geq\aq{q}{n}\geq 0\,.
\end{equation}

Now we can write
\begin{equation}\label{hp}
\dim\Mc{q+r}=\sum_{k=1}^n\,\aq{q+r}{k}=
\sum_{k=1}^n\,{r+k-1\choose r}\aq{q}{k}\,.
\end{equation}
This is the {\em\hp\/} of the \de~$\Rc{q}$ (it can be written explicitly as a
polynomial in~$r$).  Analyzing the number of arbitrary \tc s in the power
series
expansion of the general solution and comparing with these dimensions yields
the
following result.

\begin{theorem}\label{arbfun1}
The general solution of a first-order system of \de s~$\Rc{q}$ contains
$f_k$~functions depending on $k$~arguments where the numbers~$f_k$ are
determined by
\begin{equation}\label{arbaq1}
  \astretch{1.5}
  \begin{array}{c}
     f_n = \aq{1}{n} = m-\bq{1}{n}\,,\\
     f_k = \aq{1}{k}-\aq{1}{k+1} = \bq{1}{k+1}-\bq{1}{k}\,.
  \end{array}
\end{equation}
\end{theorem}

(\ref{aseq})~ensures that the~$f_k$ are always non-negative. Note that
Theorem~\ref{arbfun1} refers to algebraic representations of the general
solution, i.e.\ no integrals or derivatives of the arbitrary functions do
occur.
One can derive more general results covering also higher-order equations and
more general representations of the solution, but we will not need them here.

We define a {\em gauge symmetry\/} as a fiber-preserving transformation of the
bundle~$\E$ depending on some arbitrary functions of all independent variables
which maps solutions into solutions. (This implies that $f_n$ cannot vanish for
a system with such a symmetry.) In gauge theories one identifies solutions
related by a symmetry transformation. In order to obtain information about the
arbitrariness of the physically relevant part of the solution space we must
adjust the \cc s.

Let us assume that the gauge transformation can be written in the following
form
\begin{equation}\label{gaugerel}
\astretch{1.5}
\begin{array}{c}
\bar x^i=\Omega^i(x^j)\,,\\
\bar u^\alpha=\Lambda^\alpha\bigl(x^i,u^\beta,\lambda^{(0)}_a(x),
\partial\lambda^{(1)}_a(x),\dots,\partial^p\lambda^{(p)}_a(x)\bigr)
\end{array}
\end{equation}
where $\gamma_0$~gauge functions~$\lambda^{(0)}_a$ are entering algebraically,
$\gamma_1$~gauge functions~$\lambda^{(1)}_a$ are entering through their first
derivatives etc. Ref.~\cite{wms:diss} shows how one can handle more general
cases using a pseudogroup approach based on an implicit representation of the
transformations by \de s.

Under this assumption the gauge correction term~$\Delta\aq{q}{k}$ which must be
subtracted from $\aq{q}{k}$ to adjust for the effect of the symmetry can be
computed recursively through
\begin{equation}\label{delcc}
\Delta\aq{q}{k}=\frac{(k-1)!}{(n-1)!}\sum_{l=0}^p\gamma_l\s{n-1}{n-k-1}{q+l}-
\sum_{i=k+1}^n\frac{(k-1)!}{(i-1)!}\,\Delta\aq{q}{i}\s{i-1}{i-k}{0}\,,
\end{equation}
where the $\s{n}{k}{q}$ denote some combinatorial factors, the modified
Stirling
numbers (earlier called symmetric $q$-products) introduced in
Ref.~\cite{wms:diss,wms:str}.

\section{Constrained Dynamics \`a la Dirac}\label{dirac}

Let $q^i$ be coordinates in some $N$-dimensional configuration space~$Q$. We
restrict our exhibition to autonomous systems, as explicit time dependencies
can
always be treated by considering the time as additional coordinate in an
extended configuration space.  The dynamics of a system is then determined by
the condition that its action
\begin{equation}
S=\int L(q^i,\dot{q}^i)\,dt
\end{equation}
is stationary along trajectories~$q^i(t)$, where $L$ is the Lagrangian of the
system. It is well-known from the calculus of variations that this leads to the
Euler-Lagrange Equations
\begin{equation}\label{ele}
\frac{d}{dt}\left(\diff{L}{\dot{q}^i}\right)-\diff{L}{q^i}=0\,,
\qquad i=1,\dots,N\,.
\end{equation}

We pass from the Lagrangian formalism to the Hamiltonian one by a Legendre
transformation.  We introduce the canonically conjugate momenta
\begin{equation}
p_i=\diff{L}{\dot q^i}\,.
\end{equation}
For regular systems the Legendre transformation provides a one-to-one mapping
between the velocities~$\dot q^i$ and the momenta~$p_i$. In a constrained
system
this does no longer hold; instead one obtains by elimination some primary
constraints
\begin{equation}
\phi_\ell(q^i,p_i)=0\,.
\end{equation}
This implies that not every point of the phase space is accessible for the
system (or can be used as initial data) but only a submanifold, i.e.\ some of
the coordinates~$q^i$ do not correspond to true degrees of freedom.

The canonical Hamiltonian of the system given by
\begin{equation}\label{Hc}
H_C=p_i\dot q^i-L(q^i,\dot q^i)
\end{equation}
represents no longer the only possible choice. We can add arbitrary
combinations
of the constraints without changing its value on trajectories. This leads to
the
total Hamiltonian
\begin{equation}\label{Ht}
H_T=H_C+u^\ell\phi_\ell
\end{equation}
where the multipliers~$u^\ell$ are a priori arbitrary functions of~$q^i,p_i$.

The constraints must remain stable under the evolution of the system.
Introducing the Poisson bracket
\begin{equation}
\Bigl\{F(q^i,p_i),G(q^i,p_i)\Bigr\}=
\diff{F}{q^i}\diff{G}{p_i}-\diff{G}{q^i}\diff{F}{p_i}
\end{equation}
we can express the evolution of any observable~$F(q^i,p_i)$ concisely
\begin{equation}
\dot F=\left\{F,H_T\right\}\,.
\end{equation}
Thus we are lead to the requirement
\begin{equation}\label{seccon}
\left\{\phi_\ell,H_T\right\}\approx 0\,.
\end{equation}
The sign $\approx$ signals that this is a so-called weak equality, it must hold
only after taking all constraints into account. By a standard argument in
differential geometry this implies that the Poisson bracket in~(\ref{seccon})
must be a linear combination of the constraints. There are three possibilities
for~(\ref{seccon}): (i)~it yields modulo the constraints an equation of the
form~$1=0$; (ii)~it becomes~$0=0$; (iii)~we obtain a new equation
$\psi(q^i,p_i)=0$.

(i)~means that our equations of motion are inconsistent. This implies that they
do not possess any solution. Hence the Lagrangian is physically invalid.
(ii)~is
of course the desired outcome. (iii)~results in a secondary constraint. It is
added to the other ones. We must of course then check whether all secondary
constraints remain stable under the evolution of the system, i.e.\ we have to
repeat the procedure until we either encounter case~(i) or all constraints lead
to case~(ii). This is the so-called Dirac algorithm.

If secondary or higher constraints occur, we must distinguish whether or not
they depend on the multipliers~$u^\ell$. If yes, we can solve for some of them
which are then no longer arbitrary. This indicates the presence of second class
constraints, as a first class constraint~$\psi$ Poisson commutes weakly with
all
other constraints~$\phi_\ell$, i.e.
\begin{equation}
\left\{\psi,\phi_\ell\right\}\approx 0\,.
\end{equation}

It is well-known that first class constraints generate gauge symmetries.
Second
class constraints correspond to unphysical degrees of freedom; a typical
example is the pair $q^1=0$ and $p_1=0$. These unwanted degrees of freedom can
be eliminated using the Dirac bracket. Let $\chi_\ell$ denote all second class
constraints and define the matrix~$C$ by
\begin{equation}
C_{ij}=\left\{\chi_i,\chi_j\right\}\,.
\end{equation}
This matrix is always non-singular and we can define
\begin{equation}
\left\{f,g\right\}_*=\left\{f,g\right\}-
\left\{f,\chi_i\right\}(C^{-1})^{ij}\left\{\chi_j,g\right\}\,.
\end{equation}
In the canonical quantization of the system the Dirac brackets and not the
Poisson brackets are transformed into commutation relations.

One of the fundamental goals in constrained dynamics is to count the number of
degrees of freedom of the system. If there are $N_F$~first and $N_S$~second
class constraints in the system, then the number~$F$ of dynamical degrees of
freedom is given by
\begin{equation}\label{dfclass}
F=N-N_F-\frac{1}{2}N_S\,.
\end{equation}
This simply reflects the fact that two second class constraints are necessary
to
eliminate a degree of freedom, as we need one for the coordinate and one for
the
momentum. A first class constraint leads to a symmetry and thus to an
arbitrariness in a coordinate (or a momentum). A real elimination requires a
gauge fixing condition, i.e.\ we add a new constraint which turns the first
class constraint into a second class one.

\section{Involution Analysis}\label{point}

Now we will analyze the equation of motions from the point of view of formal
theory.  Let the bundle $\E$ be given by $\E=Q\times T$, where $Q$ is the
$N$-dimensional configuration space with coordinates $q^i$ and $T$ the time
axis, together with the natural projection $\E\rightarrow T$.
(\ref{ele})~represents a second-order equation whose symbol is determined by
the
Hessian matrix
\begin{equation}
M_{ij}=\frac{\partial^2L}{\partial\dot{q}^i\partial \dot{q}^j}\,.
\end{equation}
If the symbol has rank~$N$, the Euler-Lagrange Equations are normal and no
constraints occur. Its general solution is parameterized by $2N$~arbitrary
constants.

If, however, the symbol has lower rank, it is possible to eliminate the
second-order derivatives in some of the equations. Now it is no longer obvious,
whether or not (\ref{ele}) is involutive. Since we are dealing with an \ode,
the
symbol is always involutive. But the prolongation of the obtained differential
equations of lower order might lead to \intc s, if the resulting equations are
independent of the remaining second-order equations in~(\ref{ele}). Then we
have
to check whether some of these conditions are again of lower order; in that
case
we have to repeat the procedure.

After a finite number of iterations we will obtain either an inconsistency or
an
involutive system~$\Rcp{2}{s}$ of the following form
\begin{equation}\label{laginv}
\Rcp{2}{s}\,:\,\left\{\astretch{1.5}
\begin{array}{lll}
\ddot q^j=f^j(q^i,\dot q^i,\ddot q^n)\,,\quad&j=1,\dots,J,\ &n=J+1,\dots,N\,,\\
\dot q^k=g^k(q^i,\dot q^n)\,,&k=1,\dots,K,\ &n=K+1,\dots,N\,,\\
q^m=h^m(q^n)\,,&m=1,\dots,M,\ &n=M+1,\dots,N\,,
\end{array}\right.
\end{equation}
with $M\leq K\leq J$. Refs.~\cite{sm:dyn,sun:condyn} contain detailed
treatments
of constrained systems in the Lagrangian formalism. A closer look reveals at
once that it corresponds exactly to the completion algorithm presented in
Section~\ref{compl} applied to a system of second-order \ode s. Zeroth and
first
order equations are called there constraints of A and B type, respectively.

To relate our approach to the standard one by Dirac we pass again by a Legendre
transformation to the Hamiltonian formulation. At the level of the \de s this
means that instead of the configuration space~$Q$ the phase space~$P$ is used
to
construct~$\E$. In other words, we introduce $N$~additional dependent
variables~$p_i$ and transform~(\ref{ele}) into the first-order equation
\begin{equation}\label{ham}
\Rc{1}\,:\,\left\{
\sep{0.5}\astretch{2}
\begin{array}{lcl}
\dot{p}_i&=&\displaystyle\diff{L}{q^i}\,,\\
p_i&=&\displaystyle\diff{L}{\dot{q}^i}\,,
\end{array}\right.\qquad i=1,\dots,N\,.
\end{equation}
The first set of equations consists of course just of the Euler-Lagrange
Equations with the second set of equations plugged in.

If the matrix $M_{ij}$ has full rank, the second set of equations can be solved
for the~$\dot{q}^i$ and (\ref{ham})~is a normal equation. Otherwise we obtain
some algebraic equations of the form $\phi_\ell(q^i,p_i)=0$. They are of course
the primary constraints of Dirac. Just as in the Lagrangian formulation,
involution of~(\ref{ham}) depends on the behavior of the prolonged equations
\begin{equation}\label{procon}
D_t\phi_\ell=\diff{\phi_\ell}{q^i}\dot{q}^i+\diff{\phi_\ell}{p_i}\dot{p}_i=0\,.
\end{equation}
We are only interested in these equations restricted to~$\Rc{1}$. This
corresponds to the weak equalities used in the last section. Like in the Dirac
algorithm there are three possibilities for the result of the restriction of
each of the equations in~(\ref{procon}): (i)~it yields an inconsistency;
(ii)~it
vanishes identically; (iii)~we obtain a new independent equation.

If secondary constraints appear, we must repeat the procedure to check the
consistency of the equations of motion~(\ref{ham}). After a finite number of
steps we will either have found an inconsistency or we will have constructed an
involutive equation of the form
\begin{equation}\label{haminv}
\Rcp{1}{s}\,:\,\left\{
\sep{0.5}\astretch{1.5}
\begin{array}{ll}
\dot{p}_i=\displaystyle\diff{L}{q^i}\,,\qquad&i=1,\dots,N\,,\\
\dot{q}^j=f^j(q^i,\dot{q}^n,p_i)\,,\qquad&j=1,\dots,M\,,\ n=M+1,\dots,N\,,\\
\phi_\ell(q^i,p_i)=0\,,\qquad&\ell=1,\dots,K\,.\\
\end{array}\right.
\end{equation}

In the Dirac approach one does not use the Hamiltonian equations~(\ref{ham}),
but one introduces some multipliers and takes those derived from the total
Hamiltonian~(\ref{Ht}). To justify this we look at the differential of the
canonical Hamiltonian~(\ref{Hc})
\begin{equation}
dH_C=\dot{q}^i dp_i-\diff{L}{q^i}dq^i+
           \Bigl(p_i-\diff{L}{\dot{q}^i}\Bigr)d\dot{q}^i\,.
\end{equation}
Thus on~$\Rc{1}$ we obtain
\begin{equation}
dH_C\,\restrict{15}{\Rc{1}}=\dot{q}^i dp_i-\diff{L}{q^i}dq^i\,.
\end{equation}

Two one-forms which coincide when restricted to the constraint surface, i.e.\
the submanifold of~$\E$ defined by the constraints~$\phi_\ell=0$, can differ
only by a linear combination of the form~$u^\ell d\phi_\ell$ with arbitrary
coefficients~$u^\ell$. Since
\begin{equation}
dH_C=\diff{H_C}{q^i}dq^i+\diff{H_C}{p_i}dp_i\,,
\end{equation}
we obtain the following equations of motion (sometimes called Hamilton-Dirac
equations) living in an extended phase space
\begin{equation}\label{hamdir}
\bar\Rc{1}\,:\,\left\{\astretch{2}
\begin{array}{l}
\ds\phantom{-}\dot q^i=\diff{H_C}{p_i}+u^\ell\diff{\phi_\ell}{p_i}\,,\\
\ds -\dot p_i=\diff{H_C}{q^i}+u^\ell\diff{\phi_\ell}{q^i}\,,\\
\phi_\ell(q^i,p_i)=0\,.
\end{array}\right.
\end{equation}
Here the coefficients~$u^\ell$ must be considered as additional functions
of~$t$
or in the language of differential equations as additional dependent variables.
(\ref{ham})~is obtained, if we use the first set of equations in~(\ref{hamdir})
to express $u^\ell$ through~$q^i,\dot q^i$. Thus both systems are equivalent.

The Dirac algorithm is equivalent to the completion of system~(\ref{hamdir}).
It
requires the analysis of the prolongations of the constraints restricted
to~$\bar\Rc{1}$. They can be concisely written using Poisson brackets
\begin{equation}
\sep{0.5}\astretch{2}
\begin{array}{rl}
D_t\phi_\ell\,\restrict{15}{\bar\Rc{1}}&\ds =
\diff{\phi_\ell}{q^i}\diff{H_C}{p_i}-\diff{\phi_\ell}{p_i}\diff{H_C}{q^i}+
u^k\left(\diff{\phi_\ell}{q^i}\diff{\phi_k}{p_i}-
         \diff{\phi_\ell}{p_i}\diff{\phi_k}{q^i}\right)\\
&=\Bigl\{\phi_\ell,H_C\Bigr\}+u^k\Bigl\{\phi_\ell,\phi_k\Bigr\}\,.
\end{array}
\end{equation}

Actually, to obtain a full equivalence we should write the multiplier as
derivatives $u^\ell=\dot v^\ell$. This is only important in the case that
second
class constraints are present. Then we obtain equations determining some of
the~$u^\ell$. In principle we must prolong then these equations, too. This
unnecessary step which yields no new information can be omitted by using
derivatives. This appears also from a physical point of view somewhat more
natural, as the multipliers correspond to velocities.

\section{Counting Degrees of Freedom}\label{free}

The classical expression~(\ref{dfclass}) for the number of degrees of freedom
depends on the distinction into first and second class constraints. This
requires, however, the introduction of a Poisson structure. We will show now
that it is possible to obtain an intrinsic expression for this number without
performing such a distinction.

We will start with the Hamiltonian equations of motion~(\ref{haminv}).  In
order
to obtain the number of degrees of freedom we must count the number of
constants
necessary to characterize a physical state.  Since $\dim\je{1}=4N$, we obtain
$\dim\Rcp{1}{s}=3N-M-K$. Thus in a power series expansion of the general
solution this number of zeroth and first order coefficients can be chosen
arbitrarily.

If we identify the arbitrary functions with $\dot{q}^i$ for $i>M$, as these are
not restricted by~(\ref{haminv}), we must subtract $N-M$~constants coming from
the arbitrary functions. Thus an initial state at $t=t_0$ is specified by
$2N-K$~constants. Depending on the choice of the arbitrary functions we will,
however, obtain different values for $q^i(t_1)$ and $p_i(t_1)$ at some later
instant~$t_1$. As these correspond nevertheless, by definition, to the same
physical state, we must subtract further $N-M$~constants for the gauge
symmetry.
Thus a physical state is specified by $N+M-K$~constants and the number of
degrees of freedom~$F$ is half of this number
\begin{equation}
F=\frac{1}{2}(N+M-K)\,.
\end{equation}

We find for $\Rcp{1}{s}$ that $\bq{1}{1}=N+M$ and hence $\aq{1}{1}=N-M$.
Expressing $M$ by $N$ and $\aq{1}{1}$ and similarly $K$ by $N$, $\aq{1}{1}$ and
$\dim\Rcp{1}{s}$ yields an intrinsic expression for~$F$ independent of any
specific representation of the manifold~$\Rcp{1}{s}$
\begin{equation}\label{numfi}
F=\frac{1}{2}\,\dim\Rcp{1}{s}-\aq{1}{1}\,.
\end{equation}

If we use instead of the Hamiltonian equations of motion~(\ref{ham}) the
Euler-Lagrange Equations~(\ref{ele}), we obtain the analogue result
\begin{equation}
F=\frac{1}{2}\,\dim\Rcp{2}{s}-\aq{2}{1}\,.
\end{equation}
Both expressions yield always the same value, as we will obtain exactly the
same
dimension and \cc\ for the final involutive equation no matter whether we work
in first- or second-order because of the different dimensions of the base
spaces~\cite{pom:eins,wms:diss}.

As a by-product this implies that $J$ in (\ref{laginv}) equals $M$ in
(\ref{haminv}) and the sum of $K$ and $M$ in (\ref{laginv}) equals $K$
in~(\ref{haminv}). Thus we have the same number of constraints in both
approaches, if we omit the introduction of multipliers. This might not be very
surprising, we will, however, see later that this does no longer hold in field
theories.

We hope to study in a future paper the distinction into first and second class
constraints in more detail. Then we will also be able to discuss the relation
between these results and the classical formula~(\ref{dfclass}). For the moment
we just note that according to Theorem~\ref{arbfun1} the general solution
contains $\aq{1}{1}$~arbitrary functions. In the classical terminology this
arbitrariness stems from the gauge transformations generated by the primary
first class constraints~\cite{htt:constr,ht:quant}. Thus their number
is~$\aq{1}{1}$.

\section{Example}\label{ex}

Consider the classical problem of a particle whose movement is restricted to
the
surface of a sphere in a $D$-dimensional space but otherwise free, often also
called the rigid rotator~\cite{fh:sphere}. Without loss of generality we can
take the radius of the sphere as one and start with the Lagrangian
\begin{equation}
L(q^i,\dot q^i,\lambda,\dot\lambda)=\frac{m}{2}\dot q^2+\lambda(q^2-1)\,.
\end{equation}
($q^2=q^iq_i$, etc.) $\lambda$ is here obviously a multiplier. The canonically
conjugate momenta~$p_i,\pi$ are given by
\begin{equation}
\astretch{1.3}
\begin{array}{cl}
p_i=m\dot q_i\,,&\quad i=1,\dots,D\,,\\
\pi=0\,.
\end{array}
\end{equation}
If we introduce a further multiplier~$\mu$, we can write the total
Hamiltonian\index{Hamiltonian (total)} as
\begin{equation}
H_T=\frac{1}{2m}p^2-\lambda(q^2-1)+\mu\pi
\end{equation}

Obviously there is one primary constraint, namely $\pi=0$. The next three steps
of the Dirac algorithm lead to the constraints $q^2=1$, $pq=0$, and finally
$p^2=-2m\lambda$. It is easy to see that all these constraints are second
class\index{constraint (second class)}. The system contains thus $D-1$~degrees
of freedom which can be formally calculated by subtracting from the dimension
of
the configuration space --- $D$~coordinates~$q_i$ plus one coordinate~$\lambda$
--- half the number of second class constraints, i.e.~2.

Now we will obtain the number of degrees of freedom using a formal analysis of
the Euler-Lagrange Equations
\begin{equation}
\Rc{2}\,:\,\left\{\astretch{1.3}
\begin{array}{l}
m\ddot q_i-2\lambda q_i=0\,,\\
q^2-1=0\,.
\end{array}\right.
\end{equation}
The completion to an involutive equation requires four projections leading to
the \intc s $q\dot q=0$, $q\ddot q+\dot q^2=0$, $\dot\lambda=0$ and finally
$\ddot\lambda=0$. After some trivial manipulations we have thus the involutive
equation
\begin{equation}
\Rcp{2}{4}\,:\,\left\{\astretch{1.3}
\begin{array}{ll}
m\ddot q_i-2\lambda q_i=0\,,\quad&\ddot\lambda=0\,,\\
q\dot q=0\,,&\dot\lambda=0\,,\\
q^2-1=0\,,&m\dot q^2+2\lambda=0\,.
\end{array}\right.
\end{equation}
It is easy to see that this represents a finite type equation and thus there
are
no first class constraints. Since $\dim\Rcp{2}{4}=2D-2$, (\ref{numfi})~yields
$F=D-1$ degrees of freedom, in perfect agreement with the Dirac analysis.

Alternatively, we can analyze the Hamiltonian equations of motion
\begin{equation}
\Rc{1}\,:\,\left\{\astretch{1.3}
\begin{array}{ll}
\dot p_i-2\lambda q_i=0\,,\quad&\dot\pi-q^2+1=0\,,\\
m\dot q_i-p_i=0\,,&\pi=0\,.
\end{array}\right.
\end{equation}
Again the system becomes involutive after four projections with \intc s
$\dot\pi=0$, $qp=0$, $p^2+2m\lambda=0$, and finally $\dot\lambda=0$. This
yields
\begin{equation}
\Rcp{1}{4}\,:\,\left\{\astretch{1.3}
\begin{array}{ll}
\dot p_i-2\lambda q_i=0\,,&\dot\pi=0\,,\\
m\dot q_i-p_i=0\,,\quad&\dot\lambda=0\,,\\
\pi=0\,,&p^2+2m\lambda=0\,,\\
qp=0\,,&q^2-1=0\,.
\end{array}\right.
\end{equation}
The analysis of this equation leads of course to exactly the same number of
degrees of freedom, as $\dim\Rcp{1}{4}=\dim\Rcp{2}{4}$ and both are finite type
equations.

\section{Field Theories}\label{field}

In the section on point mechanics we studied three different ways to write the
equations of motion: the Lagrangian equations, the Hamiltonian equations
obtained from the latter one by direct application of the Legendre
transformation, and finally the Hamilton-Dirac equations where one includes
multipliers for the primary constraints. In field theories there is even more
choice, as there exist at least two different ways to perform the Legendre
transformation.

The standard way entails the explicit choice of a time variables and leads to a
non-covariant formalism. From the point of view of \de s this approach has a
further disadvantage: it is not a truly first-order formalism, as the
Hamiltonian field equations will generally still contain second-order spatial
derivatives. Thus for the application of involution theory it is probably more
appropriate to use the so-called De Donder-Weyl approach~\cite{kan:dw,rund:var}
which leads to a covariant first-order formalism.

We will therefore restrict ourselves in the sequel to the analysis of field
equations in Lagrangian formulation and leave the discussion of the Hamiltonian
approach for the future. This suffices for the purpose of this article.

Many articles on the theory of systems with constraints have the following
structure: The theoretical results are derived in the finite-dimensional case,
i.e.\ in point mechanics; the examples and applications stem, however, from
field theories. The connection is made with a remark like ``The generalization
of these results to field theories is straightforward''. But this point of view
is a bit optimistic, as a more careful discussion (see e.g.~\cite{sun:condyn})
reveals many subtle problems.

Although on the surface the main difference lies in the fact that Poisson
brackets are now computed via functional derivatives instead of partial ones,
many elementary concepts in the finite-dimensional theory become rather tricky
in an infinite-dimensional setting. For instance linear combinations must now
be
substituted by integrations. But to require that an integral vanishes is a much
less stringent condition than the vanishing of an algebraic expression and
depends decisively on the considered function space.

Similarly, inverses as they are used in the construction of Dirac brackets are
no longer uniquely defined. Often already the distinction into first and second
class constraints can be rather problematic and statements like the number of
second class constraints is always even do not make sense any more. One must
introduce the new concept of proper and improper constraints~\cite{bct:proper}.
Further problems stemming from the choice of coordinates will be discussed in
Section~\ref{freeF}.

To really solve these problem one must usually resort to fairly complicated
methods from functional analysis.  We will concentrate in this section,
however,
on another point: The naive generalization of the Dirac analysis does {\em
not\/} correspond to the completion to involution of the field equations. Thus
in general it does not suffice to prove their consistency.

Since field equations are \pde s, involution becomes a more complicated
concept.
The prolongation of lower order equations represents no longer the only way to
generate \intc s. Constraints are mostly equations of lower class. In a typical
field theory the base space~$X$ of independent variables is a $D$-dimensional
space-time. We can identify the variable~$x^D$ with time. Thus in the usual
terminology, the equations of class~$D$ are the evolutionary ones; the
remaining
ones are constraints.

The naive generalization of the Dirac algorithm prolongs all constraints only
with respect to time, as it relies solely on Poisson brackets with the
Hamiltonian. If all constraints are of class~$D-1$, this corresponds to our
approach, because it suffices to analyze the prolongations with respect to the
non-multiplicative variables and we find in this case only one, namely~$x^D$.

But now the question arises as to what happens if constraints of lower class
appear.  Then prolongations with respect to the other multiplicative variables,
i.e.\ with respect to some spatial coordinates, may lead to additional \intc s
not considered by this naive approach.

In order to exhibit this effect in ``pure form'' we begin with a highly
unphysical example without any kinetic term in the Lagrangian density. But we
will later show that this is not the important point. Consider the class of
systems described by the Lagrangian density
\begin{equation}\label{fglag}
{\cal L}[\phi,\lambda,\mu]=
\mu[\partial_x\phi-f(\phi)]+\lambda[\partial_y\phi-g(\phi)]
\end{equation}
on a three-dimensional flat space-time with coordinates~$x,y,t$. The
fields~$\lambda,\mu$ represent again multipliers, whereas $f,g$~denote fixed
but
so far arbitrary functions. Variation with respect to $\phi$ yields the
equation
\begin{equation}
\partial_x\mu+\partial_y\lambda+f^\prime(\phi)\mu+g^\prime(\phi)\lambda=0\,.
\end{equation}
More interesting equations are obtained from the multipliers. They generate
an over-determined system for~$\phi$
\begin{equation}
\astretch{1.5}
\begin{array}{l}
\partial_x\phi-f(\phi)=0\,,\\
\partial_y\phi-g(\phi)=0\,.
\end{array}
\end{equation}
Obviously this system is consistent, if and only if $f,g$ satisfy the equation
\begin{equation}\label{fgcond}
f^\prime g=fg^\prime\,.
\end{equation}
This requires that $f$ is a multiple of~$g$. It is easy to see that only under
this condition the Euler-Lagrange Equations are involutive. We must conclude
that most of the Lagrangian densities~(\ref{fglag}) are physically invalid.

Now we look at the outcome of the naive Dirac algorithm applied to this field
theory.  Obviously all three canonically conjugate momenta
$\pi_\phi,\pi_\mu,\pi_\lambda$ vanish and represent thus primary constraints.
The total Hamiltonian density is given by
\begin{equation}
{\cal H}_T=-\mu[\partial_x\phi-f(\phi)]-\lambda[\partial_y\phi-g(\phi)]
+u\pi_\phi+v\pi_\mu+w\pi_\lambda\,.
\end{equation}
This yields the following secondary constraints
\begin{equation}
\astretch{1.5}
\begin{array}{l}
\{\pi_\phi,{\cal H}_T\}=-[\partial_x\mu+\partial_y\lambda+
                          f^\prime(\phi)\mu+g^\prime(\phi)\lambda]\,,\\
\{\pi_\mu,{\cal H}_T\}=\partial_x\phi-f(\phi)\,,\\
\{\pi_\lambda,{\cal H}_T\}=\partial_y\phi-g(\phi)\,.
\end{array}
\end{equation}
Thus we obtain exactly the Euler-Lagrange Equations above. But note the crucial
difference in the further analysis. Following the naive Dirac analysis we look
only whether these equations are {\em algebraically\/} related, i.e.\ whether
one vanishes, if we take the others into account. But this does not happen
here.

Since all constraints are second class, we continue to compute the tertiary
constraints in order to fix the multipliers~$u,v,w$ introduced in the total
Hamiltonian density. They are
\begin{equation}
\astretch{1.5}
\begin{array}{l}
\partial_xv+\partial_yw+f^{\prime\prime}(\phi)\mu u+f^\prime(\phi)v+
                    g^{\prime\prime}(\phi)\lambda u+g^\prime(\phi)w=0\,,\\
\partial_xu-f^\prime(\phi)u=0\,,\\
\partial_yu-g^\prime(\phi)u=0\,.\\
\end{array}
\end{equation}
We find the interesting phenomenon that although there are six second class
constraints for three degrees of freedom $\phi,\mu,\lambda$, they do not fix
all
multipliers. $v,w$~appear only in the first equation, hence one of them can be
chosen arbitrarily. $u$~is the solution of an over-determined system which is
consistent, if $f$ and $g$ satisfy~(\ref{fgcond}).

This example shows that the real failure of this approach lies in the purely
algebraic treatment of the constraints. In point mechanics we could always
substitute one constraint in another one, if both contained the same coordinate
$q^i$ or a (time) derivative of it. This is no longer possible in field
theories, because there might be derivatives with respect to other coordinates
present. To check the consistency of such constraints requires the analysis of
\intc s.

This analysis is surely trivial in a primitive example as above. But in more
complicated cases it is rather difficult to decide when one has found all \intc
s. Assume for instance that a field theory in 3+1~dimensions leads (among
others)
to the following constraints for some field~$\phi$
\begin{equation}
\begin{array}{c}
\phi_{zz}+y\phi_{xx}=0\,,\\
\phi_{yy}=0\,.
\end{array}
\end{equation}
In this famous example due to Janet one needs five prolongations and two
projections to show that there are exactly two \intc s, namely
$\phi_{xxy}=\phi_{xxxx}=0$.  Such problems lead at the end of the last century
to the first steps towards the development of the formal theory!

One might argue that (\ref{fglag}) is a rather peculiar Lagrangian density. But
the addition of a kinetic term for $\phi$ does not really change the outcome,
although the computations are slightly more complicated due to the appearance
of
further \intc s. Take
\begin{equation}
\bar{\cal L}=\frac{1}{2}(\partial_t\phi)^2+{\cal L}\,.
\end{equation}
Its Euler-Lagrange Equations are
\begin{equation}
\astretch{1.5}
\begin{array}{l}
\partial_{tt}\phi+\partial_x\mu+\partial_y\lambda+
                  f^\prime(\phi)\mu+g^\prime(\phi)\lambda=0\,,\\
\partial_x\phi-f(\phi)=0\,,\\
\partial_y\phi-g(\phi)=0\,.
\end{array}
\end{equation}

Only the first equation has changed. But since the only second-order derivative
involves~$\phi$, our completion procedure generates besides (\ref{fgcond}) two
further \intc s involving second-order derivatives of~$\mu,\lambda$. We omit
them here, because they are rather complicated.

We also do not show the Dirac analysis of this system. It suffices to note that
the two new conditions (in Hamiltonian form) are also found by the Dirac
algorithm. But this was to be expected, because their construction requires
prolongations with respect to time. Only (\ref{fgcond}) is again overlooked,
for
no spatial cross-derivatives are performed.

The natural question is whether systems of this form are unphysical for some
reason or whether this effect occurs often and might lead to a wrong results.
Such purely spatial cross-derivatives (or to be more precise: linear
combinations of spatial prolongations) are only necessary, if the field
equations form an over-determined system. This might appear at first sight
quite
unusual but one can find such systems in the literature (see
e.g.~\cite{got:conj}). A typical example is a gauge fixing condition built into
the Lagrangian density.

\section{Higher-Order Lagrangians}\label{hol}

Theories described by Lagrangian containing higher order
derivatives~\cite{dhot:elast,ht:crps,nest:high,pons:ostro} can be treated in
exactly the same way. Using again the notation of Section~\ref{inv} such a
Lagrangian is a function~$L(x^i,u^\alpha,\pam)$. The~$x^i$ are coordinates on
the underlying space-time in the case of a field theory or just the time in
point mechanics. Similarly, the~$u^\alpha$ denote either the fields in the
theory or the generalized coordinates and the~$\pam$ represent the derivatives.

{}From the calculus of variations it is well-known that the Euler-Lagrange
Equations can now be written using the Euler operators~\cite{olv:lgde}
\begin{equation}\label{hoele}
E_\alpha(L)=\diff{L}{u^\alpha}+
            \sum_\mu(-1)^{|\mu|}D_\mu\left(\diff{L}{\pam}\right)=0\,.
\end{equation}
If $\mu=[\mu_1,\dots,\mu_n]$, then $D_\mu=D_1^{\mu_1}\cdots D_n^{\mu_n}$ where
$D_i$ denotes the total derivative with respect to~$x^i$. Obviously we recover
(\ref{ele}), if there is only one~$x^i$, namely the time~$t$, and $L$ depends
only on derivatives with $|\mu|=1$. The sum in (\ref{hoele}) is always finite,
as $L$ contains only derivatives up to a given order.

As in the standard case there exist at least three possibilities for the
starting point of the involution analysis. The simplest choice is of course to
use directly the Euler-Lagrange Equations~(\ref{hoele}). Alternatively one can
pass to a Hamiltonian formulation. This requires now the introduction of
several
momenta conjugate to each field.  Then one can either transform the
Euler-Lagrange Equations directly or one can derive in the same manner as
before
the Hamilton-Dirac Equations by introducing multipliers.

There is no need to repeat the arguments of Section~\ref{point}, as they still
apply in the same way. Instead we consider as an example Podolsky's generalized
electrodynamics~\cite{pod:edyn} in the Lagrangian formalism.  It demonstrates
the typical way an involution proof proceeds for field theories.
Refs.~\cite{wms:diss,wms:sym} contain further examples like Yang-Mills or
Einstein Equations. The Lagrangian density is given by
\begin{equation}
{\cal L}=-\frac{1}{4}F_{\mu\nu}F^{\mu\nu}-a^2\partial_\rho
F^{\sigma\rho}\partial_\tau F_\sigma^\tau\,.
\end{equation}
The space-time indices run from 1 to~$D$; we identify $x^D$ with the time.
$a$~is a constant. If it vanishes, we recover the standard Maxwell theory.

In terms of the vector potential~$A_\mu$ the Euler-Lagrange
Equations~(\ref{hoele}) give a fourth-order system
\begin{equation}
\Rc{4}\,:\,\quad
(1-2a^2\dalembert)\dalembert A_\mu-
\partial_\mu\left[(1-2a^2\dalembert)\partial^\nu A_\nu\right]=0\,,
\quad\mu=1,\dots,D
\end{equation}
with the D'Alembertian $\dalembert=\eta^{\mu\nu}\partial_\mu\partial_\nu$.

According to Definition~\ref{invol} we must first check whether the
symbol~$\Mc{4}$ is involutive. It is given by the equations
\begin{equation}
\Mc{4}\,:\,\quad
\eta^{\mu\nu}\eta^{\rho\sigma}v^\alpha_{\mu\nu\rho\sigma}-
\eta^{o\tau}\eta^{\alpha\beta}v^\delta_{o\tau\beta\delta}=0\,,
\quad\alpha=1,\dots,D
\end{equation}
where $v^\alpha_{\mu\nu\rho\sigma}$ is a placeholder for the derivative
$\partial_{\mu\nu\rho\sigma}A^\alpha$. It is easy to see that for $\alpha\neq
D$
we can choose in each equation the variable~$v^\alpha_{DDDD}$ as pivot, i.e.\
all these equations are of class~$D$. For $\alpha=D$, however, the
corresponding
variable cancels. We can obtain at most a pivot of class~$D-1$,
e.g.~$v^D_{D-1,D-1,D-1,D-1}$. Thus
\begin{equation}\label{betaP}
\bq{4}{D}=D-1\,,\quad\bq{4}{D-1}=1\,,\quad\bq{4}{k}=0\,.
\end{equation}

One can prove that this cancellation for $\alpha=D$ does not simply stem from a
singular coordinate system either by using the tableau of the system
(cf.~\cite{wms:diss}) or by arguing that $\bq{4}{D}=D$ is not possible, as the
system has a gauge symmetry~\cite{gp:pod}. This argument relies on the results
of Ref.~\cite{wms:str} on the arbitrariness of the general solution.

In order to apply Definition~\ref{invol} we must compute next the rank of the
prolonged symbol~$\Mc{5}$. It is defined by
\begin{equation}
\Mc{5}\,:\,\quad
\eta^{\mu\nu}\eta^{\rho\sigma}v^\alpha_{\mu\nu\rho\sigma\gamma}-
\eta^{o\tau}\eta^{\alpha\beta}v^\delta_{o\tau\beta\gamma\delta}=0\,,
\quad\alpha,\gamma=1,\dots,D\,.
\end{equation}
It is easy to see that these equations are not all independent, because if we
set $\alpha=\gamma$ and sum the result vanishes.

It follows from the discussion before Definition~\ref{invol} and the obtained
values for the $\bq{4}{k}$ that $\rank\Mc{5}\geq D^2-1$. But since we have
already found an identity, the same value represents an upper bound for the
rank. Thus the rank must be exactly $D^2-1$ and by (\ref{invsymb}) $\Mc{4}$ is
involutive.

In order to see whether \intc s occur we must check whether this identity holds
only at the level of the prolonged symbol or also if we use the full prolonged
equations. We know already that the fifth order derivatives cancel, but it
could
happen that some lower order equation remains. It is, however, easy to see that
this is not the case. This is analogue to the Noether identity in the Maxwell
theory. Hence $\Rc{4}$ is involutive. This implies that in the Lagrangian
formalism only one constraint appears, namely the equation for $\alpha=D$ which
is of lower class.

For later use we note the \cc s of the theory
\begin{equation}
\aq{4}{4}=1\,,\quad\aq{4}{3}=15\,,\quad\aq{4}{2}=40\,,\quad\aq{4}{1}=80\,.
\end{equation}
Adjusting for the symmetry $A_\mu\rightarrow A_\mu+\partial_\mu\lambda$
with~(\ref{delcc}) yields the following gauge corrected values ($\gamma_1=1$)
\begin{equation}\label{gccp}
\bar\aq{4}{4}=0\,,\quad\bar\aq{4}{3}=10\,,\quad\bar\aq{4}{2}=25\,,\quad
\bar\aq{4}{1}=45\,.
\end{equation}

The Hamiltonian treatment is similar but more involved, as now secondary
constraints appear. There is no need to detail it here. Since the only
constraint in the Lagrangian formalism is of class~$D-1$, the Dirac analysis is
sufficient and equivalent to the involution analysis. Thus we just recover the
calculations presented in Ref.~\cite{gp:pod}.

The constraint analysis of Podolsky's generalized theory is very similar to the
standard Maxwell theory. In both cases we find that the Euler-Lagrange
Equations
are already involutive due to the Noether identity, whereas in the Hamiltonian
formalism we must perform a few steps until we reach an involutive system. The
same effect can be observed in other field theories. This seems to imply that
at
least at classical level the Lagrangian formalism is more efficient, as it
yields faster an involutive system.

\section{``Field Theoretical'' Degrees of Freedom}\label{freeF}

The classical procedure to count degrees of freedom in field theories is simply
to stick to the rule~(\ref{dfclass}) used in the finite-dimensional case.
$N$~denotes now the number of fields. The argument is essentially the same:
Each
constraint ``fixes'' one field in the phase space and in the case of a first
class constraint the symmetry eliminates a further degree of freedom.

The problem with this approach is that in field theories constraints are
usually
\de s. Hence they cannot really fix a field; there remains some freedom.
Speaking about arbitrary functions in the general solution the idea behind the
rule above seems to be that a degree of freedom corresponds to the possibility
to prescribe as initial data a function of $D-1$~arguments, i.e.\ of all
spatial
variables. This is for instance the case in a regular theory, as its field
equations form a normal system satisfying the conditions of the
\cako~\cite{ch:mp2}.

A generalization of this idea was already introduced by Einstein~\cite{ein:rel}
with his definition of the strength of a system of \de s. Up to a numerical
factor depending on the dimension of space-time it can be identified with the
number of arbitrary functions of~$D-1$~arguments~\cite{wms:str} provided the
system is absolutely compatible. Latter condition entails that there are no
arbitrary functions of $D$~arguments.

Of course such considerations make sense only after taking gauge symmetries
into
account, because such a symmetry leads to arbitrary functions of all
independent
variables. Einstein's definition of the strength contains such correction
terms.
Based on the formal theory of \pde s we propose as an intrinsic definition for
the number of ``field theoretical'' degrees of freedom the gauge corrected
\cc~$\bar\aq{q}{D-1}$. This definition covers also theories derived from
Lagrangians containing higher-order derivatives.

According to Theorem~\ref{arbfun1} $\bar\aq{q}{D-1}$ corresponds to the number
of arbitrary functions of $D-1$~arguments in the general solution modulo the
gauge symmetry. Of course we assume here that $\bar\aq{q}{D}=0$ as in
Einstein's
approach. Otherwise either the field equations are under-determined or the
gauge
group was not correctly identified. For a regular theory no gauge correction is
needed. In this case $\aq{q}{D}=0$ and $\aq{q}{D-1}=m$ where $m$ is the number
of fields. Thus we recover the usual result.

For the Maxwell theory one obtains the following gauge corrected
\cc s~\cite{wms:diss}
\begin{equation}
\bar\aq{2}{4}=0\,,\quad\bar\aq{2}{3}=4\,,\quad\bar\aq{2}{2}=6\,,
\quad\bar\aq{2}{1}=0\,.
\end{equation}
(The same values can be obtained without gauge correction by directly analyzing
the field equations in field strength formulation.) Thus we obtain in perfect
agreement with the usual result 4~degrees of freedom. In contrast Podolsky's
generalized theory possesses 10~degrees of freedom as evident
from~(\ref{gccp}).

It is important here to note that the \cc s are intrinsically defined and thus
independent of any specific coordinate system. This is of course a property one
should expect of a reasonable definition for the number of degrees of freedom.
As Steinhardt~\cite{stein:probl} showed with some explicit examples (see also
the discussion in~\cite{sun:condyn}), the classical approach encounters
problems, if ``wrong'' coordinates are used.

He considered among others the simple example of a free massive scalar field in
1+1~dimensions. In standard coordinates this system described by the Lagrangian
density $2{\cal L}=-\partial^\mu\phi\partial_\mu\phi-m^2\phi^2$ is obviously
regular and contains one degree of freedom. In light-cone coordinates
$x^\pm=(x\pm t)/2$ the Lagrangian density becomes
\begin{equation}
{\cal L}=\partial_+\phi\partial_-\phi-m^2\phi^2/2\,.
\end{equation}
If we choose $x^+$ as new evolution parameter, the canonically conjugate
momentum is $\pi=\partial_-\phi$ and independent of the
velocity~$\partial_+\phi$.  Hence the system is constrained.

It is quite subtle to decide whether this constraint is first or second class,
but here we are not concerned with these difficulties. The important fact is
that there appears a constraint and hence according to~(\ref{dfclass}) the
system has less than one degree of freedom! Obviously this cannot be correct.
We
are not aware of any proposal in the literature for a modified formula to count
degrees of freedom that takes this effect into account.

In the involution analysis a similar phenomenon occurs. As already Steinhardt
pointed out, the appearance of the constraint is intimately connected with the
fact that the light-cone coordinates are the characteristics of the field
equations.  Actually this represents a special case of a more general problem,
namely the \Dreg\ of a coordinate system~\cite{pom:eins} which was already
mentioned in Section~\ref{inv}.

If a coordinate system is not \dreg, the procedure to compute the $\bq{q}{k}$
described in Section~\ref{inv} yields too small values. This corresponds to the
too many constraints found by Steinhardt. For instance in our case the field
equations are
\begin{equation}
2\partial_+\partial_-\phi+m^2\phi=0\,.
\end{equation}
Obviously there is no derivative of class~2 in this equation. But it can be
generated with a simple coordinate transformation, namely going back to the
original ones. Thus the correct value for $\bq{2}{2}$ is one and not zero. This
yields $\aq{2}{2}=0$ and $\aq{2}{1}=1$. Since no gauge correction is necessary
here, we obtain the expected result: one degree of freedom.

There exists a simple method to determine the correct values of the $\bq{q}{k}$
in {\em any\/} coordinate system without performing a coordinate change. It
makes use of the generalized tableaux of a differential equation.  Their ranks
provide an intrinsic definition for the~$\bq{q}{k}$; i.e.\ their determination
is a simple problem in linear algebra. For lack of space we cannot detail this
approach here but refer to the literature~\cite{wms:tab,wms:diss}.  The
important point is that our definition for the number of degrees of freedom can
be applied to any system in any coordinate system and leads always to the same
number.

\section{Conclusion}

Obviously it is one of the most elementary requirements on a system of \de s to
be consistent, i.e.\ to possess at least a formal \pss. The formal theory of
\de
s provides us with a powerful tool to check this property: the involution
analysis.  We have shown how it can be used in a physical context, namely in
constrained dynamics. Here the \de s arise typically as the equation of motions
derived from some Lagrangian~$L$.

One should keep in mind that the motivation for the Dirac analysis is exactly
the same. Its first task is less to exhibit all constraints but to prove the
consistency of the equations of motion. Hence it is not very surprising that we
find that for finite-dimensional systems the Dirac analysis coincides with the
involution analysis. Both approaches yield (in some sense merely as a
by-product) all constraints or \intc s, respectively, of the system.

Although it seems to be a commonly accepted claim that the Dirac analysis can
be
extended without modifications to field theories, we have given examples where
in our opinion this classical approach is not sufficient. Their construction
was
based on the simple observation that Dirac takes only the temporal evolution of
the system into account. He does not consider spatial prolongations. Hence his
approach is incomplete and not able to prove the consistency of the field
equations.

One cannot really speak of a failure of the Dirac approach, but one must note
that it must be augmented by some kind of analysis of the spatial \intc s. One
could for instance use an approach like the involution analysis for this
purpose.  But we think that it is conceptually easier to use one method instead
of a combination of several ones.

We believe that the involution approach is more flexible than the Dirac method.
It can be applied to the Lagrangian as well as to the Hamiltonian formalism. In
a future article we will show that the so-called symplectic method for
first-order Lagrangian~\cite{fj:ham} also fits into this scheme. The same holds
for higher-order Lagrangians. As soon as the equations of motion are obtained,
either by a variational principle or some other way, the involution analysis
can
start. This is especially important for systems with anholonomic constraints
which cannot be treated with the usual variational methods~\cite{rund:var}.

An important advantage of the involution analysis is that it represents a
geometric framework, i.e.\ all constructions are intrinsic and coordinate
independent. This is not true for the Dirac approach. It is well-known that the
number of constraints can be different in different coordinate systems. A
typical example are light-cone coordinates. In contrast, our definition of the
number of degrees of freedom for a field theory via the \cc s is completely
intrinsic. To our knowledge no such definition has so far been proposed in the
literature.

One might wonder where all the subtleties of the Dirac analysis like regularity
conditions on the constraints, ineffective or reducible constraints have
disappeared~\cite{ht:quant}. They are of course still present. Most of them are
hidden behind the calculation of the dimension of the various
submanifolds~$\Rcpp{q}{r}{s}$ used in the completion algorithm depicted in
Fig.~\ref{ckad}. But this is a classical problem in geometry and for special
types of constraints there may exist alternative approaches. For instance
polynomial non-linearities represent probably the most important case in
applications.  For them we can avoid a discussion of most of the mentioned
problems by using Gr\"obner bases techniques~\cite{kw:dim}.

Another effect which might lead to problems is that the rank of the symbol (or
more generally the numbers~$\bq{q}{k}$) does not need to be constant. It might
change, if certain additional \de s hold. This generalizes the classification
of
the eigenvalues of the Hessian of the Lagrangian as given by
Lusanna~\cite{lus:noe}.

Finally, one should note that by making contact with the formal theory of \de s
one obtains suddenly a well understood object, namely an involutive system.
Many
properties of such systems are known and many techniques have been developed
for
their further analysis. All these results are now available for constrained
systems.

\section*{Acknowledgments}

We thank M.~Henneaux for helpful comments. {\small WMS}~is supported by a grant
of the School of Physics and Materials, Lancaster University and {\small
RWT}~is
grateful for support by a grant in the {\small EC} Human Capital and Mobility
program.

\end{document}